\documentclass[pre,twocolumn,superscriptaddress]{revtex4-1}

\usepackage{dcolumn}
\usepackage{amsmath}
\usepackage{graphicx}
\usepackage{calrsfs}
\usepackage{bm}
\usepackage[table,xcdraw]{xcolor}
\usepackage{enumitem}
\usepackage{hyperref}
\usepackage{cancel}

\begin{document}
\title{Homologous nodes in annotated complex networks}

\author{Sung Soo Moon}
  \affiliation{Department of Chemical Engineering and Biotechnology, \\University of Cambridge, Philippa Fawcett Drive, Cambridge CB3 0AS, UK}

\author{Sebastian E. Ahnert}%
  \affiliation{Department of Chemical Engineering and Biotechnology, \\University of Cambridge, Philippa Fawcett Drive, Cambridge CB3 0AS, UK}
    \affiliation{The Alan Turing Institute, 96 Euston Road, London NW1 2DB, UK}

\begin{abstract}
Many real-world networks have associated metadata that assigns categorical labels to nodes. Analysis of these annotations can complement the topological analysis of complex networks. Annotated networks have typically been used to evaluate community detection approaches. Here, we introduce an approach that combines the quantitative analysis of annotations and network structure, which groups nodes according to similar distributions of node annotations in their neighbourhoods. Importantly the nodes that are grouped together, which we call {\em homologues} may not be connected to each other at all. By applying our approach to three very different real-world networks we show that these groupings identify common functional roles and properties of nodes in the network. 
\end{abstract}
\maketitle

\section{Introduction}
\label{sec:intro}
Many complex systems can be represented as networks and studied using network analysis. This has given rise to the field of network science over the past decades \cite{newman_networks_2018}. One aim of network science has been to derive interpretable descriptors of nodes, edges, and substructures in the network from the network connectivity. However, in many real networks nodes and edges are not homogeneous. Various levels of abstraction exist for node and edge features that add further layers of information to the network connectivity. For instance in network neuroscience, weighted and directed edges, or signed edges can represent the density of synapses between brain regions \cite{bassett_network_2017}, edge types can specify neurotransmitters \cite{lin_network_2024} and node labels can annotate physical positions or cell classifications \cite{hobert_revisiting_2016}. How these annotations supplement our analysis and understanding of the structural organisation of the network is an open problem. 

A common algorithm employed in the analysis of complex networks is community detection, which typically partitions a network such that nodes are more densely connected within each community than between different communities \cite{fortunato_20_2022}. For instance, stochastic block modelling approaches \cite{newman_structure_2016} have been effective at detecting communities by inspecting the density of connectivity between distinct partitions of the network. Community detection has been applied successfully in many different real-world networks, and recent work has generalised this concept\cite{newman_generalized_2015} and explored hierarchical communities \citep{schaub_hierarchical_2023}. Many current approaches that aim to identify the roles that individual nodes play in a network use metrics of modular organisation. An example is the participation coefficient \cite{guimera_cartography_2005, guimera_functional_2005, guimera_classes_2007, pedersen_reducing_2020} which characterises nodes that facilitate communication within, and between modules. However, sparsity of some real networks, such as disease networks, can make the detection and functional identification of disease modules difficult using community detection \citep{sharma_disease_2015, ghiassian_disease_2015}. While modularity may suffice to identify nodes with similar annotations in systems whose annotation schemes are homophilic or assortative (i.e. two nodes of same type more likely to be connected \cite{newman_assortative_2002, noldus_assortativity_2015, rizi_homophily_2024, newman_mixing_2003}), a framework for characterising nodes with complex connectivity profiles is under-explored.

Much of the prior work on the categorisation of nodes has focussed on metrics that describe a node’s position in context of the whole network. Centrality measures have been employed to provide a ranking of most important nodes in the global structure of the network, and thus disparate parts of the system can be meaningfully compared \cite{koschutzki_centrality_2008, das_study_2018}. A wealth of social network literature on these measures aims to classify node importance and to relate topological features to social roles defined extrinsically \citep{sun_predicting_2016}. This is often also framed as a node classification problem \citep{bhagat_node_2011}, where features of the network are used for inferring or propagating predefined annotations for incomplete data \citep{newman_network_2018} or in order to extrapolate annotations to another dataset \citep{henderson_rolx_2012}. The literature on Graph Representation Learning \cite{tenorio_structure-guided_2024, corso_graph_2024} also describes signals in local network structure that relate to meaningful groupings of nodes, and that can be used to infer annotations. 

Here, we introduce an approach to characterise nodes by the annotations of nodes in their local neighbourhood. More specifically we construct interpretable feature vectors from these annotations and use them to cluster nodes. We describe nodes that contain similar annotations in their network neighbourhood as {\em homologous}. This terminology is a generalisation of the concept of `serial homologues' in neuronal networks (see Discussion), which describe neurons with similar connectivity neighbourhoods. 

We propose to group nodes that have analogous, or homologous, local neighbourhoods but that are not necessarily form modular structures, meaning we are able to group nodes that are not necessarily homophilic or assortative, (i.e. share edges with each other). Instead, homologues display similar sets of neighbours in terms of the neighbours' annotations. Our method can be applied to undirected, directed, and bipartite networks, and can be extended to weighted networks. We demonstrate that the node classes identified by our approach have interpretable functional properties, and thus provide evidence of network homologues in a variety of real-world networks: a gene regulatory network, a food web, and a recipe network.


\section{Results}
\begin{figure*}
\includegraphics[width=0.95\textwidth]{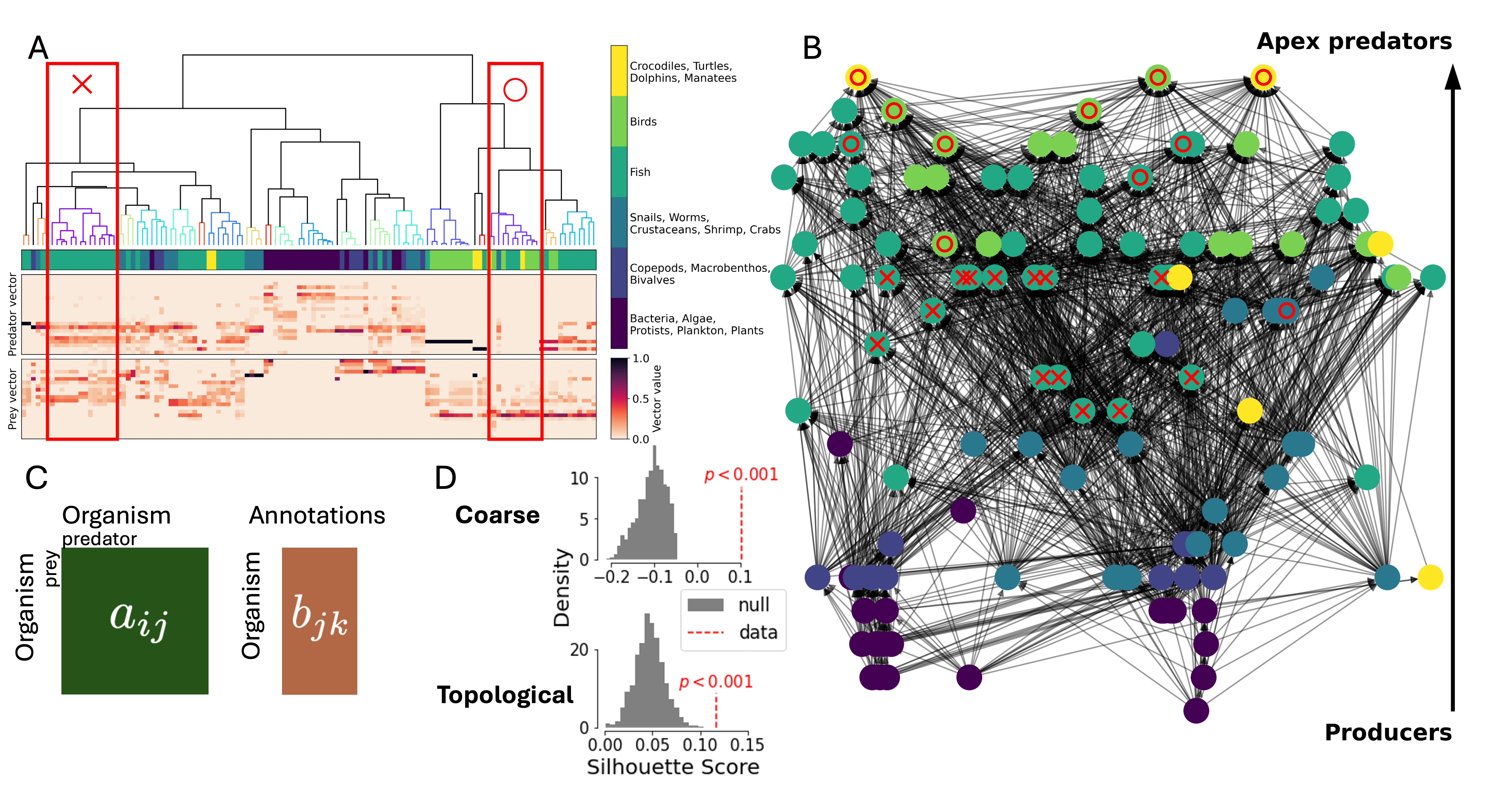}
\caption{\label{fig:fw1} Food web clustering and visualisation. \textbf{A}, Hierarchical clustering with organism types binned into 6 coarse categories (see colour bar). We highlight the two largest clusters ($\times$ and $\bigcirc$ in red) when the dendrogram is cut at $0.2 \times$ the maximum height. The heatmap below the dendrogram shows that the clusters are organisms grouped by their similarities of the annotations in their local in-neighbourhoods (predators) and out-neighbourhoods (prey), thus inferring organisms that play similar roles in the food web. Note that cluster $\times$ of homologous organisms is homogenous in its composition, consisting entirely of fish, whereas cluster $\bigcirc$ is a heterogeneous group of different organisms. \textbf{B}, Hierarchical layout of the food web, where a directed edge from $a$ to $b$ denotes the flow of biomass (i.e. $a$ is eaten by $b$). The bottom nodes are producers, and apex predators are at the top. Labelled in red $\times$ and $\bigcirc$ are the same clusters as before, which reveals that the heterogenous group $\bigcirc$ represents apex predators, and the homogenous group $\times$ represents a group of fish that are all located at a similar level in the food web hierarchy. \textbf{C}, Illustration of the food web adjacency matrix $a_{ij}$ and the organism annotations adjacency matrix $b_{jk}$. To create a null distribution, the columns of $b_{jk}$ are randomly permuted. \textbf{D}, The silhouette score measures how well the clusters align with the 6 coarsest organism annotations shown in \textbf{A}, and the topological annotations from stratifying the food web (see Methods). The grey distribution is generated using the null model described above, using 1000 randomisations. The dotted red lines show the silhouette score of the observed network. In both cases the observed data deviates from the null distribution with a $p$-value of $p < 0.001$.}
\end{figure*}

\subsection{Predator and prey profiles of organisms in a food web reveal diverse sets of organisms with similar ecological roles}

A food web is a network that describes the  predator-prey relationships between a set of organisms in an ecosystem. We model these relationships as an unweighted, directed graph, in which nodes represent the organisms and directed edges leading from prey to predator specify the predator-prey relationships. The top (apex) predators without any predators themselves have no outgoing edges, and at the bottom of the food chain micro-organisms and plants that do not prey on other organisms have no incoming edges.

There are both coarse and fine levels of annotations available regarding the organism type for every node. We use the finest level to construct the neighbourhood annotation vectors. Incoming and outgoing connections are separately aggregated and normalised, thus characterising each node in terms of its combined predator and prey profile. We then apply hierarchical clustering to group organisms that are similar in terms of these vectors (see figure \ref{fig:fw1}A). The clustering reveals groupings of organisms that are (a) not necessarily connected to each other, and (b) not necessarily similar in terms of organism type, but which tend to be on a similar hierarchical level of the food chain and thus play a similar role in the ecosystem (figure \ref{fig:fw1}A, B and Supplementary Information for a fully annotated figure). We cut the dendrogram at $0.2\times$ the total height and inspected the clusters. The top and bottom level organisms are easily distinguished by their absence of out- or in-vectors respectively. The clusters containing the intermediate-level organisms form several classes of organisms that are situated at similar levels in the food web and often (but not always) are of similar organism types. We highlight the two largest clusters in the dendrogram (labelled $\times$ and $\bigcirc$ in red), and show their positions in the network (see Figure \ref{fig:fw1}A and B). The $\times$ cluster is a homogenous cluster of fish in the middle of the food web, while the $\bigcirc$ is a diverse cluster of top predators. 

To quantify the the extent to which clusters group organisms of a similar type we use the silhouette score to determine the alignment between embeddings of the node (i.e. pairwise vector distances) and the annotations of the nodes themselves. Specifically, the score calculates how close in distance the organisms we know to belong to the same group are compared to other organism groups. A silhouette score is calculated for each ``grouping'' of nodes, then averaged to give a total score between -1 to 1, where 1 is complete alignment of the embedding and the grouping, and -1 is misalignment, and values around 0 indicate overlapping clustering (see Methods for details). 

Using this measure, we tested for statistical significance of the silhouette score of the observed network compared to label-permuted null models (see figure \ref{fig:fw1}C). The null models were constructed by shuffling labels and recalculating connectivity vectors, and with each instance, a silhouette score is evaluated to build the null model distribution. We tested for statistical significance with various levels of organism categorisation in the food chain from coarse to fine. For the coarse organism categorisation, we find that our silhouette score is positive ($0.1$) and statistically significant ($p<0.001$), thus organisms within these categories have similar predator and prey profiles (figure \ref{fig:fw1}D, top). We tested whether finer organism types can also be retrieved, and while statistically significant against the null, the embedding did not correlate overall to these labels (see Supplementary Information). We also considered the topological ordering (see Methods) that stratifies the hierarchy of the network, asking the question: is the observed node labelling predictive of the topological network structure or does the structure constrain the homologous behaviour? We find that the topological labels are aligned with the vector embedding with a positive silhouette score ($s=0.12$) that is statistically significant, thus exhibits homologous relationships (figure \ref{fig:fw1}D, bottom).

\subsection{Recipe network vectorisation}
\subsubsection{Clustering reveals sets of ingredients that are characteristic for particular cuisines}
Next, we apply our approach to the recipe network: a tripartite network of recipes,  ingredients, and cuisine annotations. We first project this multipartite network by counting the number of recipes in each cuisine associated with each ingredient, and create a cuisine-space embedding vector for each ingredient. In other words, a given ingredient will be represented by a vector in which each entry signifies a given cuisine's proportion of recipes that include that ingredient. Mathematically, let $a_{ij} = 1$ be the bipartite graph adjacency matrix entry of recipe $j$ being associated with cuisine $i$, and zero otherwise. 
Also, let $b_{jk}$ be the bipartite graph adjacency matrix element where $b_{jk}=1$ if ingredient $k$ is used in recipe $j$, and $b_{jk}=0$ otherwise. Then, the standardised number of times an ingredient $k$ is used in a particular cuisine $i$ is $v_{ki} = \sum_{j} a_{ij}b_{jk} /\sum_{j'}a_{ij'}$, where the denominator corrects for the number of total recipes in each cuisine. 
Thus, the normalised contribution of cuisine $k$ for ingredient $i$ is $p_{ki} = v_{ki}/\sum_{i}v_{ki}$. 

\begin{figure*}
\includegraphics[width=0.95\textwidth]{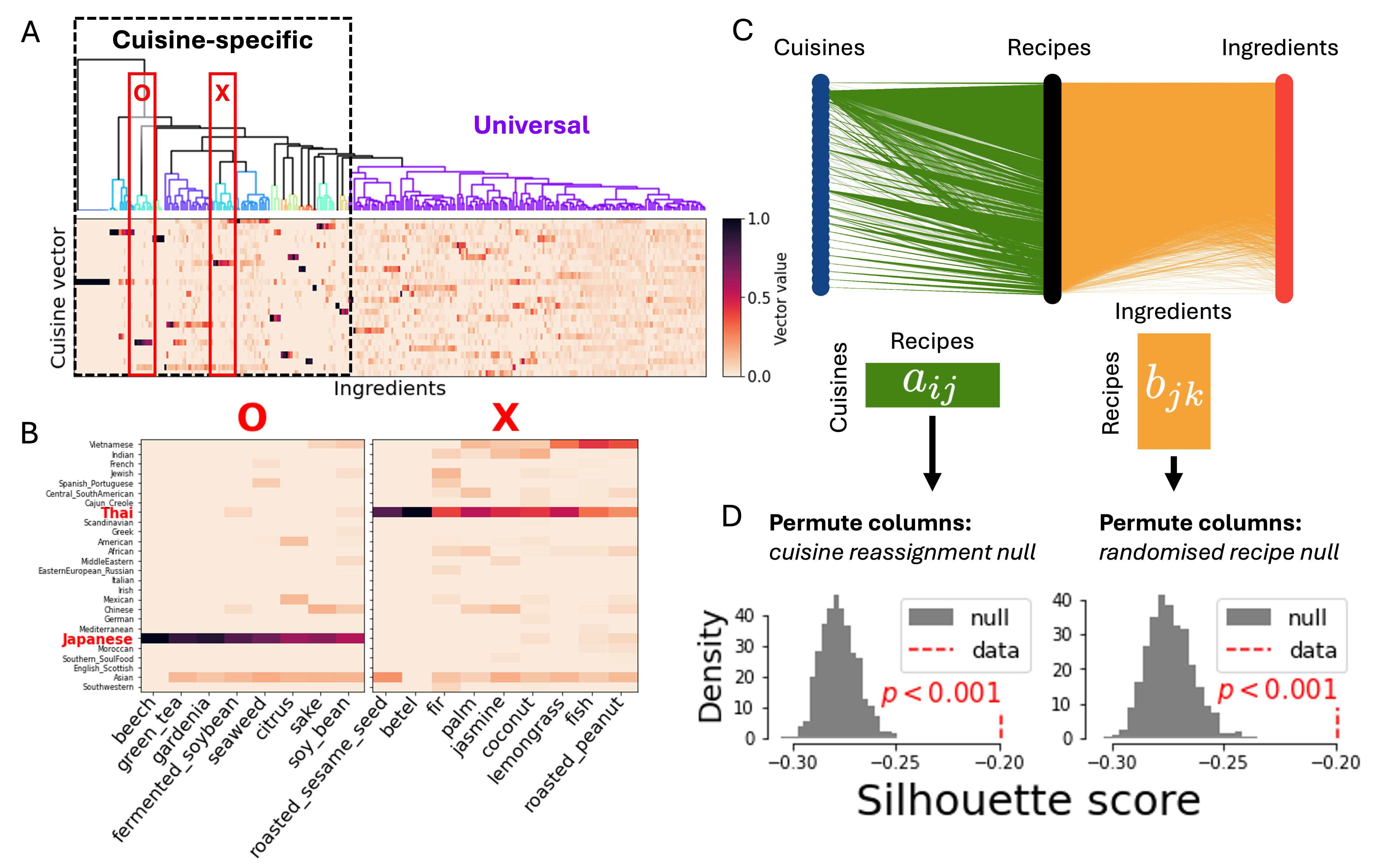}
\caption{\label{fig:rn1} \textbf{A}, Hierarchical clustering of the ingredients in the recipe network using the cuisine-space vectors. The tree is cut at $0.3\times$ the height of the dendrogram to categorise the ingredients into those that are highly cuisine-specific and those that are universal. In \textbf{B} we demonstrate two cuisine-specific clusters. One is a strongly \textbf{Japanese}  set of ingredients (marked \textbf{O} in red) and the other is strongly associated with \textbf{Thai} cuisine (marked \textbf{X} in red). In \textbf{C} we illustrate two null models. If we rewire the cuisine-recipe bipartite layer, we effectively permute the columns of the adjacency matrix, to give the cuisine reassignment null. If we rewire the recipe-ingredient bipartite layer, we effectively permute the columns of the recipe-ingredient adjacency matrix giving the randomised recipe null. \textbf{D}, We test how aligned the vector embedding is with respect to ingredient type by using the silhouette score, which was negative indicating that ingredients of the same type were not generally clustered together $s=-0.20$. Here, the dotted red lines mark the values of the observed silhouette scores and the associated $p$ values were estimated from 1000 null model realisations. In both cases the silhouette score of the observed data deviates significantly from the null model distributions, with p-values $< 0.001$.}
\end{figure*}

When these vectors are clustered, they reveal groupings of ingredients that display similar relative usage across different cuisines (figure \ref{fig:rn1}A). The clusters, therefore, are ingredients that have a similar role in each cuisine, since they have similar usage across all cuisines. Broadly, we identified two meta-categories of clusters: cuisine-specific and universal, which were identified by cutting the tree at $0.3\times$ the height of the dendrogram. We identified two cuisine-specific clusters for inspection in figure \ref{fig:rn1}A and B, shown labelled in red \textbf{O} and \textbf{X}. These clusters are a group of ingredients strongly associated with \textbf{Japanese} cuisine and \textbf{Thai} cuisines respectively, as shown in the heatmap of the cuisine-space vectors. 

We test the ingredient clustering with the silhouette score against the ingredient type. The silhouette score is negative ($s=-0.20$), suggesting that ingredient combinations that are particular to cuisines or sets of cuisines are heterogeneous. For instance, the two clusters \textbf{O} and \textbf{X} we have highlighted are heterogeneous groups of ingredients of different types. 

To test whether the observed network and annotations were correlated by chance, we construct two different null models: a permutation test that randomly reassigns cuisine membership of recipes, and a permutation test using randomised recipes (see figure \ref{fig:rn1}C, D and Methods). We find that the observed values differ from the null models in a statistical significant way, which indicates that the distribution of ingredient types of the observed clusters is very unlikely to arise by chance. This is despite the fact that the fat-tailed distribution of the recipe association of ingredients \citep{ahn_flavor_2011} will inevitably give rise to some clusters by chance in any null model realisation.  

\begin{figure}
\includegraphics[width=\linewidth]{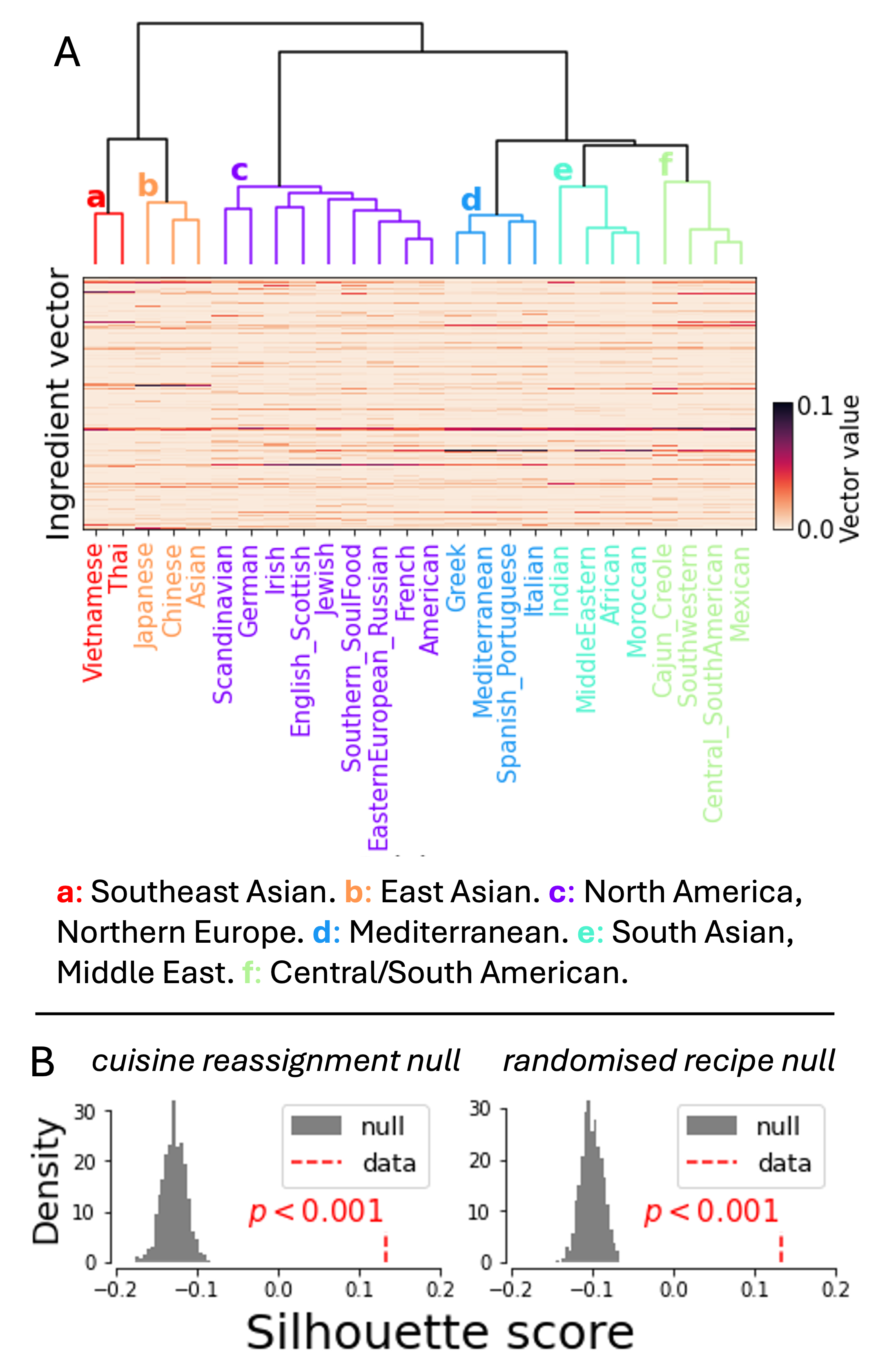}
\caption{\label{fig:rn2} \textbf{A}, Hierarchical clustering of cuisines in the recipe network using ingredient-space vectors. The tree is cut at $0.4\times$ the height of the dendrogram to show six clusters of cuisines that arise from their similarity in terms of the ingredients used across the recipes in a given cuisine (labelled in coloured boxes). The categorisations are consistent with geographical and historical canon of culinary culture. In \textbf{B} we show the null distributions of the silhouette score, evaluated the clustering with respect to geographical cuisine origin. The silhouette score positive indicating good agreement with geographical labelling ($s=0.13$). The dotted red lines mark the values of the observed silhouette scores and the associated $p$ values were estimated from 1000 null realisations. In both cases the silhouette score of the observed data deviates significantly from the null model distributions, with $p$-values $< 0.001$.}
\end{figure}

\subsubsection{Clustering reveals hierarchical similarity of cuisine types}

In a parallel process to the construction of cuisine-space vectors of ingredients, we now assemble the ingredient-space vectors of cuisines in the following way: Using the bipartite graph matrix elements $a_{ij}$ and $b_{jk}$ as before, the number of times an ingredient $k$ is used in a given cuisine $i$ is $\nu_{ik} = \sum_{j}a_{ij}b_{jk}$. The normalised contribution of ingredient $k$ in cuisine therefore becomes $\rho_{ik} = \nu_{ik}/\sum_{k'}\nu_{ik'}$. This defines our probability vectors for hierarchical clustering (figure \ref{fig:rn2}A). We broadly see three strands of cuisine clusters that join together - Southeast and East Asian, North American and Northern European, and the third cluster could further be split into three: Southern European and South American, with Middle Eastern and South Asian, in line with previous studies on recipe networks \cite{sharma_hierarchical_2020}. 

The order in which cuisines hierarchically form clusters are consistent historically and geographically, the latter of which we test with the silhouette score, by measuring the extent to which the embedding aligns with cuisines grouped by their geographical region (see Methods). We find that the embeddings correlate to the geographical grouping ($s=0.13$), and this is statistically significant from the two permutation null models we tested (see figure \ref{fig:rn2}B).

\subsection{Downstream connectivity profiles can be used to classify transcription factors in a gene regulatory network}
We next apply our approach to the \emph{Arabidopsis thaliana} gene regulatory network, with an associated bipartite metadata graph of Gene Ontology (GO) annotations \citep{berardini_functional_2004, badia-i-mompel_gene_2023, ahnert_form_2016, taylor-teeples_arabidopsis_2015}. The regulatory network is directed and unweighted, and the GO terms and the genes or gene products they are associated with form a bipartite graph. Here, each node is assigned a distribution of discrete labels (associated GO terms), but the connectivity vectorisation process generalises intuitively. For every gene, the GO term frequency vector is calculated by summing up the GO term assignment in the GO bipartite network. Next, in the regulatory network of genes: a particular gene has in-edges if they are regulated by transcription factors or out-edges if they themselves are transcription factors regulating other genes. So, with the target gene's downstream and upstream connections, the in and out connections can be aggregated by the GO term vectors associated with these partners. We ignore any GO terms that appear only once in the network, since we are not able to draw meaningful conclusions regarding these terms through clustering. Our resulting vectors are normalised by dividing each GO term contribution by the sum of the in or out GO term vector (see Methods).

\begin{figure*}
\centering
\includegraphics[width=0.9\textwidth]{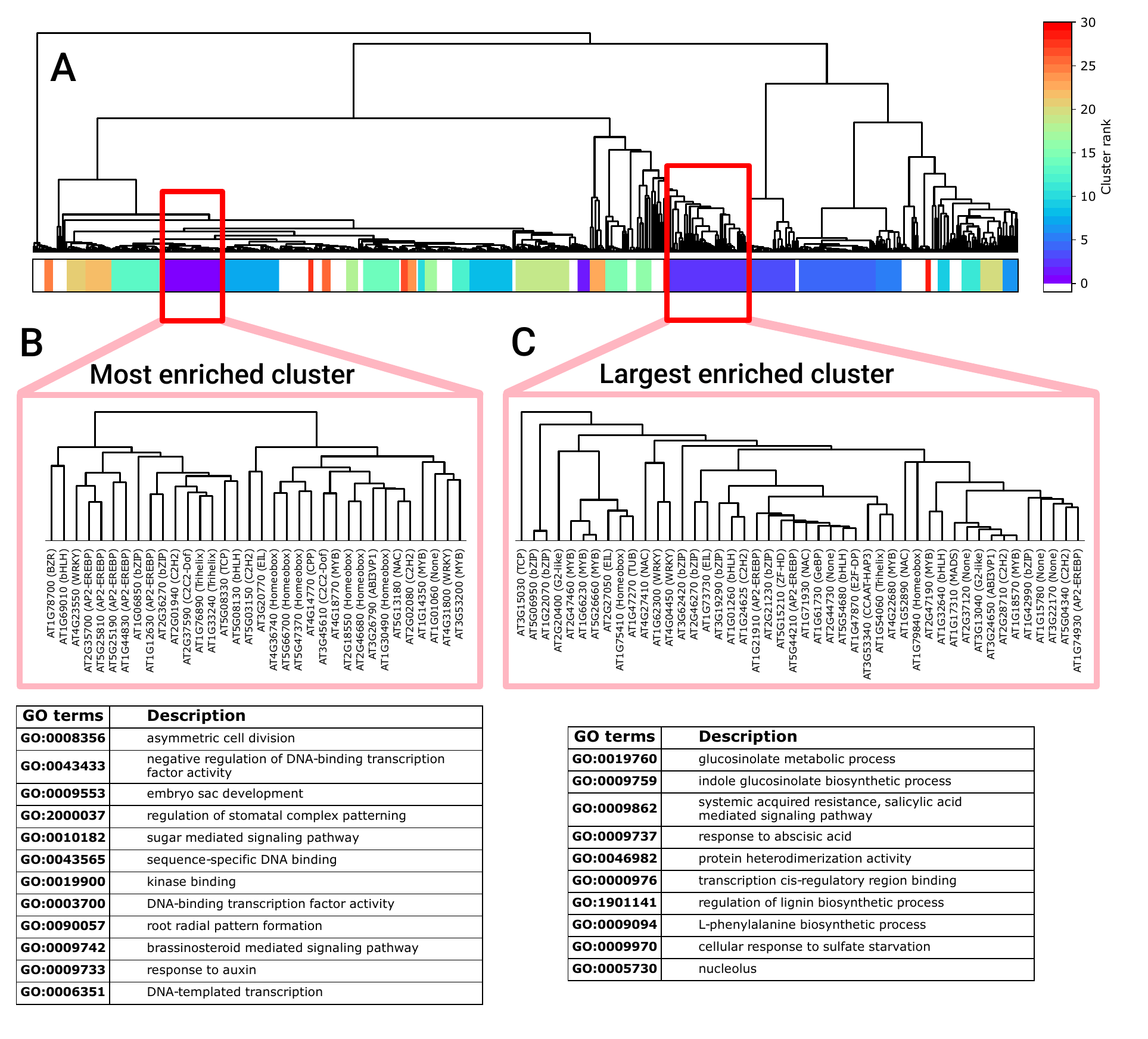}
\caption{\label{fig:grn1} Clustering of the gene regulatory network using vectors derived from Gene Ontology (GO) labels. \textbf{A} shows the dendrogram, with coloured labels indicating membership in enriched clusters sorted by the number of enriched GO terms (see main text). White signifies no enrichment. We highlight the cluster with the most enriched terms (\textbf{B}) and the largest enriched cluster (\textbf{C}). The leaves are labelled with the gene ID and transcription factor family. Additionally, the enriched GO terms and their descriptors are provided for the two clusters visualised, which are sorted from most enriched to least. For the full analysis and data of the TF clusters, labels and their enriched GO terms see the Supplementary Information. 
}
\end{figure*}

Following the connectivity vectorisation procedure, we construct the incoming and outgoing connectivity annotation vectors, with each node associated with a distribution of GO term labels that are aggregated via the regulatory network of TFs and genes. Using hierarchical clustering, we cluster the TFs by the connectivity profile of the GO terms of their ingoing and outcoming neighbours (figure \ref{fig:grn1}A). We use an enrichment analysis (with the Benjamini-Hochberg procedure to control for false discovery rate, $\alpha=0.1$) to extract the clusters that are highly enriched in multiple specific GO terms to determine whether the clusters of genes are characterised by shared functions \cite{zhou_bipartite_2007} (see Methods). We ranked all possible clusters produced by the dendrogram, in other words all possible branch points of the tree, first by the number of enriched GO terms (priority given to larger number of enriched terms), and second by the size of the cluster (priority to larger). Then, we discarded any clusters that shared subsets of TFs with a higher-ranked cluster. The final result of this procedure is a non-overlapping set of clusters with enriched GO terms, (figure \ref{fig:grn1}A). These clusters have associated GO terms that are overrepresented by chance, and thus statistically significant from the pool of TFs. We are able to assign 559 of 580 TFs ($96\%$) into one of thee functionally enriched clusters.

The clusters reveal the extent of to which the nodes in the local downstream neighbourhood of TFs are functionally similar (see Supplementary Information for full clustering and associated GO terms). Such enriched clusters could be used to supplement incomplete metadata on TFs and Targets, if genes with unknown function exhibit similar connectivity to downstream targets as known TFs, or are connected upstream to TFs that have Targets with a particular functional profile according to our approach.


\section{Discussion}
In the neuroscience literature, there exists a thread of work on \emph{connectivity fingerprints} or \emph{profiles} of neurons that are characteristic of specific brain regions, and which enable comparison across individuals and species. Although related to our paper, this prior work focuses on macroscopic connectivity which relies on highly unique landmarks in the annotations of coarse brain regions \cite{mars_connectivity_2013, mars_connectivity_2018, mars_whole_2018, passingham_anatomical_2002}. Additionally, the interplay between biological attributes of nodes and their structural connections has been studied in conventional assortative-disassortative frameworks \citep{bazinet_assortative_2023}. However, for fine-grained annotations, we require an extension to existing frameworks. Recently, in neuron-resolution connectomics (structural connectivity maps of the brain), many studies have categorised cells into groups of neurons that are connected in the same way, often constituting a cell type \cite{bazinet_towards_2023, hobert_revisiting_2016, jonas_automatic_2015, schlegel_whole-brain_2024, scheffer_connectome_2020}. 
For example in the fruit fly ventral nerve cord, there exist six leg neuropils (major regions dense with synapses) that are modular structures controlling each leg \citep{marin_systematic_2024}. The structure is such that there are identical copies of neurons that have the analogous connectivity (and proposed functional) role within their local module, but are not strongly connected to each other. Instead, they share an upstream controlling circuitry that facilitates coordination. These copies of neurons that follow the symmetry in different modular components are called serial homologue groups \citep{marin_systematic_2024}. These homologues have been characterised by their local connectivity patterns, as their identical developmental genetic code generates analogous connectivity within each leg neuropil. Identifying intrinsic organisational principles comes at a great benefit of studying the developmental, anatomical and also functional aspects of the system.

Our use of the term `homology' is inspired by this similarity of connective context. In some cases our usage overlaps more concretely with the concept of homology in biology, which implies a common evolutionary origin \citep{haszprunar_types_1992, shepherd_postembryonic_2016, jensen_orthologs_2001}. In all three datasets studied in this paper it is possible to identify common origins for some of the homologues in the network. Importantly however this analogy does not hold in general, as can be seen in the case of the apex predators in the food web, and also by the fact that in many real-world networks originate in ways that do not resemble any kind of evolutionary process.

We find that clustering the local connectivity profiles of the nodes uncovers deeper correlations and relationships than simple disassortative community structure the current language of network science can express. How should we interpret these clusters? In the recipe network, cuisine phylogeny is recovered. In the food web, we clustered organisms by their predator and prey profiles, therefore correlating organism type and topological position to the network structure. In the gene regulatory network similar groups of transcription factors in their biological function can be discovered by their connectivity profiles. In analogy with the previous examples, clustered TFs have similar GO term profiles of their neighbours. This consolidates that the connectivity profile of the TF and its own function are correlated, and we are able to group TFs by their higher level family characterization, while collectively the subgraph of the regulatory network they inhabit are very sparsely connected. These relationships are not an explicit part of the network structure themselves as cuisines share no edges with each other, and organisms in the same strata tend not to predate in direct loops. Rather, we uncover a developmental process of the network; the clustering structure of cuisines embodies remarkable parallels to culinary culture and history, and biological networks are directly formed from evolutionary processes. The results suggest that annotations themselves relate to a functional or explainable quality of the network structure, and that we have statistically demonstrated the uniqueness of the observed network with the observed annotations as one synergistic system. For these reasons, analogous components in disparate communities or modules are ``homologous'' in our description of the term.

\section{Conclusion}

Our approach combines network topology with node annotations to identify `homologues', meaning nodes that display similar distributions of annotations in their directed or undirected network neighbourhoods. Our applications demonstrate the advantages of this approach in example networks of various types, and a variety of annotations. We provide functional interpretations of homologous sets of nodes by examining whether these groupings correlate with metadata properties or structural characteristics. This approach may also be valuable in the context of incomplete metadata, as putative node labels could be inferred from homologues. The generality of this method makes it applicable to any annotated network, whether it is directed or undirected, weighted or unweighted, unipartite or multipartite. Due to this fact, and the complementarity to existing community detection approaches, we believe that this method provides a valuable new way to connect topology and function in annotated networks.

\section{\label{sec:methods}Methods}
\subsection{Connectivity vectorisation}
We first describe the simplest case of an undirected, unweighted network with $N$ nodes and $M$ different annotation types. Let $\textbf{A}^{(N \times N)}$ be an adjacency matrix where element $a_{ij} = 1$ if an edge exists between node $i$ and $j$ and zero otherwise. We define the annotation scheme for nodes as a bipartite graph adjacency matrix $\textbf{B}^{(N \times M)}$ where $b_{jk}=1$ if node $j$ has label $k$ and zero otherwise. For a given target node with index $i$, we can aggregate all its connectivity via the labelling scheme of its neighbours by performing the matrix multiplication, and so the component contributing to the $k^{\text{th}}$ label of the connectivity vector for node $i$ becomes:
\begin{equation}
v_{ik} = \sum_{j}a_{ij}b_{jk} = (\textbf{A} \cdot \textbf{B})_{ik}.
\end{equation}

To normalise the vectors for each node, we simply divide by the sum of total connectivity. In an unweighted network, this is their degree. Then, the probability that node $i$ connects to a node with annotation type $k$ becomes:
\begin{equation}
p_{ik} = \frac{ v_{ik}}{\sum_{k'} v_{ik'}}.
\end{equation}

To extend this method for directed, weighted graphs, we simply substitute in the directed, weighted adjacency matrix. Element $a_{ij}$ is now the weighted directed edge from node $i$ to $j$, and the bipartite graph adjacency remains the same. Here, we must calculate in and out probability vectors separately. To aggregate the directed connectivity by node annotations is as before. Firstly, the out-vector is identically written:
\begin{equation}\label{eqn:out_vec}
v_{ik}^{\text{out}} = \sum_{j}a_{ij}b_{jk} = (\textbf{A} \cdot \textbf{B})_{ik} ; \qquad p_{ik}^{\text{out}} = \frac{v_{ik}^{\text{out}}}{\sum_{k'} v_{ik'}^{\text{out}}},
\end{equation}
and the in-vector is the same, with a transposed adjacency matrix: 
\begin{equation}\label{eqn:in_vec}
v_{ik}^{\text{in}} = \sum_{j}a_{ji}b_{jk} = (\textbf{A}^T \cdot \textbf{B})_{ik} ; \qquad p_{ik}^{\text{in}} = \frac{v_{ik}^{\text{in}}}{\sum_{k'} v_{ik'}^{\text{in}}}.
\end{equation}

Another natural extension to the formulation is that the bipartite labelling matrix need not be a one-to-one labelling. Suppose that nodes can have multiple labels associated to it, and so $b_{jk} = 1$ for all labels $k$ associated to node $j$. Now, following the same procedure as before will also produce connectivity vectors calculated in the same way as equations \ref{eqn:out_vec} and \ref{eqn:in_vec}, with the interpretation that it jointly pools all possible contributions from the annotations as if each node with more than one annotation were multiple nodes for each annotation contributing to the connectivity. With this, the method is versatile and can be applied to a combination of (un)weighted, (un)directed networks with structure that can be expressed as an adjacency matrix. 

We exploit the fact that the connectivity aggregation is nothing but simple matrix multiplications - so  we use the sparse matrices data type implemented in \texttt{scipy.sparse} for a fast and memory efficient implementation.

We note that each node will have a uniquely calculable connectivity vector for which we can use clustering tools to extract relationships between them. We use hierarchical clustering with the Euclidean metric and Ward's method implementations in \texttt{scipy.cluster.hierarchy}. 

\subsection{Silhouette score}
The silhouette score or coefficient is calculated from a distance matrix between different nodes in their normalised connectivity vectors. Let $d_{ij}$ be the distance from node $i$'s vector to node $j$, and let the set $G_I$ be the group membership of node $i$ where $i\in G_I$. Then we define two quantities:
\begin{equation}
a_i = \frac{1}{|G_I| - 1} \sum_{j\in G_I, i\neq j}d_{ij},
\end{equation}
which is the mean distance from node $i$ to all the other nodes in group $I$, and
\begin{equation}
b_i = \min_{J\neq I} \frac{1}{|G_J|} \sum_{j\in G_J}d_{ij},
\end{equation}
which is the average distance to the closest group $J$ distinct from $I$. Then the silhouette coefficient for $i$ is:
\begin{equation}
s_i = \frac{b_i - a_i}{\max (a_i, b_i)}.
\end{equation}
This measures how well node $i$ fits with the rest of the nodes in group $I$ in the presence of closest proximity group. We take the average silhouette score over all nodes to give a total score. We use the implementation of \texttt{silhouette\_samples} from the \texttt{sklearn} package in \texttt{python}.

\subsection{Null models for annotation connectivity clustering}
For the gene regulatory network, we employ an ontology enrichment analysis for a statistical test, detailed in the next subsection. To test for statistical significance of our clustering results in the food web and recipe networks, we construct null models for testing how deviant the observed network is from a random distribution. The randomisation procedure is to shuffle node annotations and recalculate the probability vectors. Then, by inspecting the pairwise distances of the connectivity vectors and using the silhouette score for observed and null networks, we can estimate the probability of the observed network within our null model. In our models we obtain 1000 realizations of randomised null silhouette score distributions to estimate the $p$ value for the observed network. Quoted $p$ values arise from probability in the null distribution to get a value as extreme as the observed silhouette score. If no such null realisation was greater than or equal to the observed, we estimate the $p$ value to be bounded from above by the probability of the most extreme value, which would be $1/1000$.

The food web network consists of organisms as nodes with organism categories as annotations used to create annotation connectivity vector profiles $v_{ik}$ giving the incoming and outgoing contribution of organism type $k$ for node $i$. The null connectivity vectors $v_{ik}'$, are constructed from randomly shuffling the annotations. Then, we can compare the categorisations in two ways. The first is the identity consistent method where the shuffled annotations redefine the node identities we group to evaluate the newly calculated vectors. This tests how the observed network deviates from a random assignment of node identities and asks whether random chance can correlate the node identity to the profile of connectivity to annotations. In the food web network we shuffle the organism types and newly relabelled nodes retain those types when evaluating the embedding performance using the silhouette score. Alternatively, we can perform an identity preserving null model where we group the shuffled vectors by their fixed node identities. This tests whether the network structure lends itself to similarities in the connectivity profiles. An example of this would be a topological ordering of nodes which is a property of the network structure rather than the annotations, thus these identities are invariant under label permutations.  

The recipe data is a tripartite network where recipes have associated cuisine type and a list of ingredients constituting them. Here there are two choices of relabelling: we can shuffle the recipes by their cuisines type or shuffle ingredient membership of each recipe. When we shuffle the recipes by their cuisine type, it preserves individual recipes’ constituent ingredients, identical to sampling a set of real recipes and assigning a cuisine label at random. When we shuffle the ingredient membership, we are creating random recipes but keeping fixed the number of ingredients in each recipe. We can apply these two nulls to the two different clusterings we perform. Firstly, clustering ingredient-space cuisine vectors to reveal similar cuisines by their similar use of ingredients. Secondly, clustering cuisine-space ingredient vectors to reveal similar ingredients that have been utilised in similar ways among cuisines of the world. To test for statistical significance in the cuisine clustering, we compare the distance matrix of cuisines to geographical regions of cuisines in the silhouette score calculations. To statistically test for significance in ingredient embedding, we consider ingredients by type (manually labelled).


\subsection{Gene Ontology Enrichment Analysis}
The Gene Ontology (GO) labels \citep{boyle_gotermfinderopen_2004, ashburner_gene_2000, berardini_functional_2004} allow us to test our vector clustering for statistical significance. We aggregate multiple hypothesis tests using the hypergeometric distribution of finding each of the GO term frequencies for a given cluster of genes, correcting for false discovery rate using the Benjamini-Hochberg method \cite{boyle_gotermfinderopen_2004}. 

Consider a hypergeometric probability distribution to model the probability of finding $k$ genes associated with GO term $g$ from a sample of $n$, where the total population is $N$ and the total number of genes with GO term $g$ in the population is $K$. The probability is then: 
\begin{equation}\label{eqn:hypergeo}
    p(X = k) = \frac{\binom{K}{k} \binom{N-K}{n-k}}{\binom{N}{n}},
\end{equation}
where $\binom{a}{b}$ is shorthand for the binomial coefficient: $\frac{a!}{b!(a-b)!}$. 

We are interested in GO terms present in the cluster with at least 2 genes that are associated with it. This is to filter out the many genes that are only seen once in the cluster as we cannot draw useful comparisons with it across multiple genes. For each GO term, we consider its prevalence in the cluster of genes, we calculate the combined probability of observing $k$ or more genes in that cluster of size $n$:
\begin{equation}\label{eqn:p_val}
    p(k \leq X \leq n) = \sum_{i=k}^{n}p(X=i),
\end{equation}
giving us the $p$-value for a one-tailed test. 

The $p$-value for each of the GO term of interest is extracted. As the probability of a rare event increases with the number of hypotheses tested, we can correct the false discovery rate using the Benjamini-Hochberg method. Firstly, our desired aggregate significance level over all hypotheses is set to $\alpha$. Then, we rank the $p$-values of the individual hypotheses $p_1, p_2, ..., p_m$ for testing GO terms $g_1, g_2, ..., g_m$, where $m$ is the number of hypotheses, and $p_i < p_j \text{ for } i < j$. We pick the largest $j$ for which $p_j \leq \frac{j}{m}\alpha$. All null hypotheses $0, 1, ..., j$ are rejected as significant over-representation compared to a random sample, therefore GO terms $g_1, g_2, ..., g_j$ are functionally significant for the set of genes under investigation. 

Using this method we can determine the GO terms that are statistically significant and over represented and discard the ones that are within the significance level of a random, independent sample. We set the $\alpha$ level to to 0.1 for our analyses. 

\subsection{Probabilistic graph traversal}
To topologically sort the food web, we use a probabilistic traversal model since the network is not strictly acyclic to calculate a unique absolute order structure. In a directed, weighted graph $G$, we start with a set of nodes to begin the traversal through the graph. The model assigns a probability of of a node $j$ being picked at the next from node $i$: 
\begin{equation}\label{eqn:layer_model_2}
    p_{ij}= \text{min}\bigg(\frac{w_{ij}}{c*\sum_k w_{ik}} , 1 \bigg),
\end{equation}
where $w_{ij}$ is the weight of directed edge $(i,j)$ and $c$ is some corrective factor \citep{cheong_transforming_2024, schlegel_information_2021}, but here we use $c=1$. We start our traversal from the bottom feeders as seed nodes and determine the mean path length (over 1000 realisations, and rounded to the nearest integer) as the `layer' stratifying the food web network. 

With the permuted annotation vectors in the null model, we calculate the silhouette score each time. The observed network and labels had a silhouette score of $s=0.12$, which indicates some agreement between the embedding and the topological layer label. The null model distribution of silhouette scores was also positive but their agreement with the labelling was significantly weaker ($p<0.001$). The topological layer testing is an example of the identity preserving null as the topological layer is a property of the network and not the labels, so were preserved in the null calculations. 

\subsection{Data and code}
We used the Florida food web data \cite{heymans_network_2002}, \emph{Arabidopsis thalania} gene regulatory network and associated GO annotations \citep{berardini_functional_2004, badia-i-mompel_gene_2023, ahnert_form_2016, taylor-teeples_arabidopsis_2015}, and the allrecipes recipe network \cite{ahn_flavor_2011}. All raw and derivative data is available from \url{https://doi.org/10.5281/zenodo.15525130}. All code used to run the analysis and generate figures is available from \url{https://github.com/sungsushi/vectorisation}.

\section*{Funding}
This work was supported by Harding Distinguished Postgraduate Scholars Programme Leverage Scheme and the Engineering and Physical Sciences Research Council Doctoral Training Partnerships [EP/T517847/1] (both to S.S.M.). 

\begin{acknowledgments}
We thank Gregory S. X. E. Jefferis, Philipp Schlegel, Tomke St\"{u}rner and Elizabeth Marin for helpful discussions. 
\end{acknowledgments}

\bibliography{v_bib}

\appendix
\renewcommand{\thefigure}{A\arabic{figure}}
\setcounter{figure}{0}

\section{Full labelled figures of connectivity vector clustering}
All hierarchical clustering dendrograms were created with the Euclidean metric and Ward's method. Full resolution images are also available as part of the Supplementary Materials, and in the repository \url{https://github.com/sungsushi/vectorisation}.

\subsection{Food web network}
\begin{figure*}
\includegraphics[width=\textwidth]{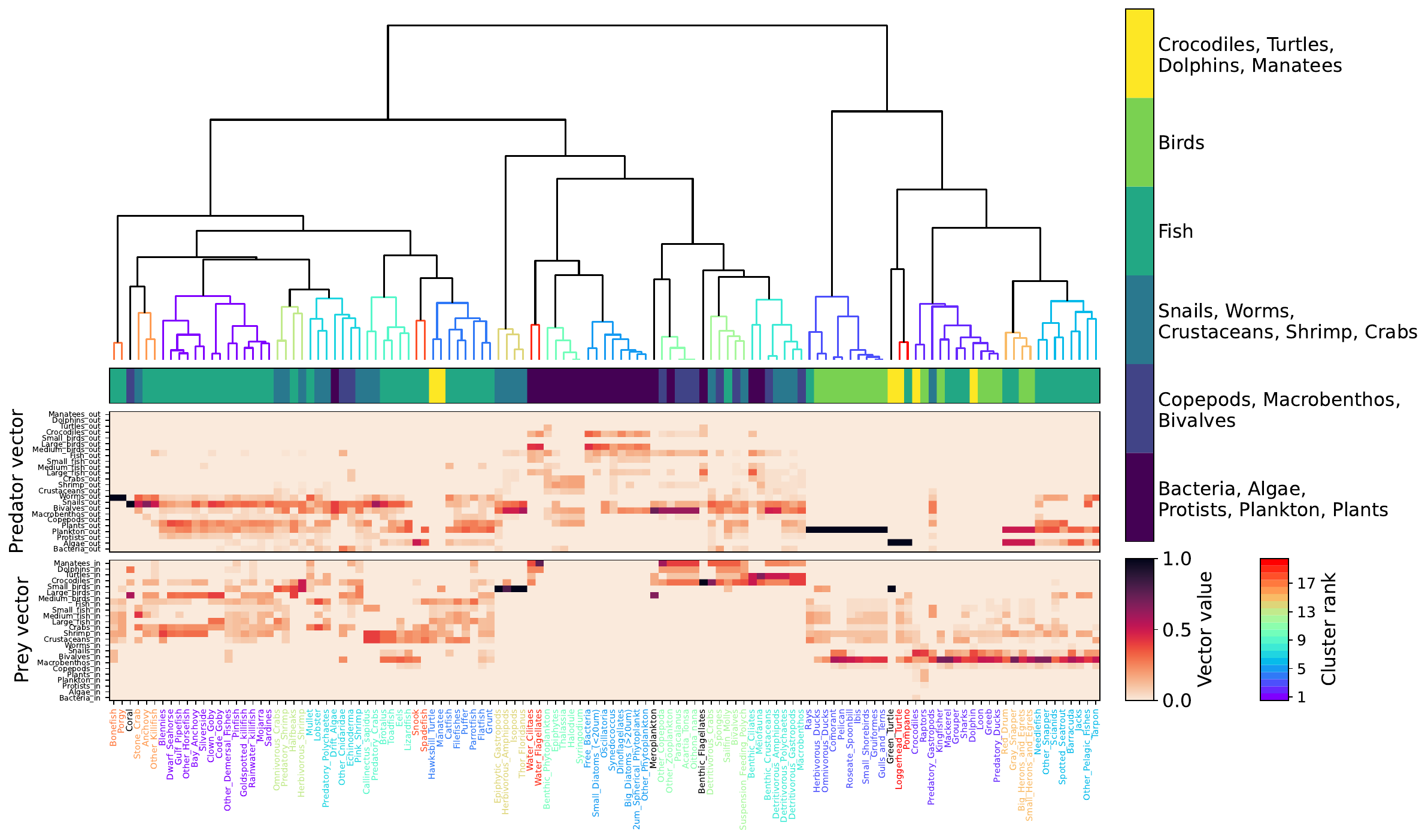}
\centering
\caption{\label{fig:s2:fw}\textbf{Hierarchical clustering with of the food web network.} Here, the dendrogram is clustered by the predator and prey vector profiles (out-vector and in-vectors respectively). All leaves from the dendrogram are labelled. The shown clusters arise from the dendrogram cut at $0.2\times$ the dendrogram height, ranked by size.}
\end{figure*}

In the food web network, the nodes represent organisms and the edges denote the direction of biomass transfer. Thus, aggregating the edges by incoming and outgoing directions groups a given organism's prey and predators respectively. This is the basis of the connectivity vectors. We have used the finest animal categorisations for the vectorisation process. Hierarchical clustering is applied to these vectors, and the fully labelled dendrogram, and their associated vectors as heatmaps are shown in figure \ref{fig:s2:fw}.

\subsection{Recipe network}
The recipe network is a multipartite network of recipes, belonging to cuisine labels, and also associated to constituent ingredients. We may use a bipartite projection to build vectors of cuisines by their normalised frequency of recipes using each ingredient (we call ingredient-space cuisine vectors in the main paper), or, build vectors of ingredients by their normalised frequency of cuisines that utilise them. Since there are an unequal sample of recipes of each cuisine, we have corrected for this (see Methods). Hierarchical clustering is applied to these vectors. Here, we present the fully annotated figure of dendorgram clustering the cuisine-space vectors of ingredients, and their associated vectors as heatmaps in figure \ref{fig:s2:fw}.

\begin{figure*}
\includegraphics[width=1\textwidth]{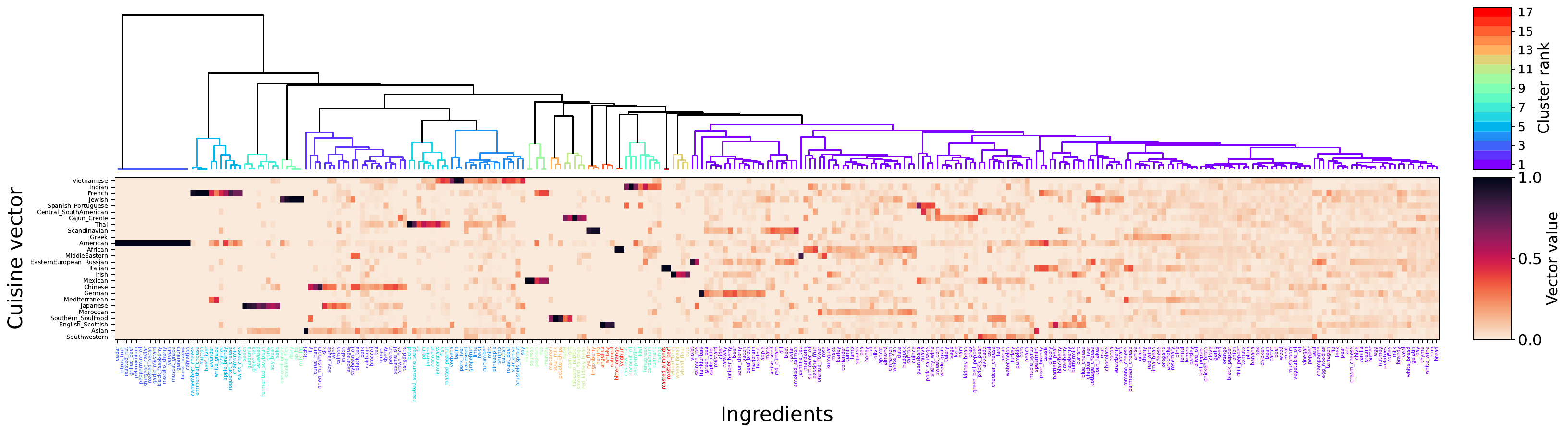}
\centering
\caption{\label{fig:s3:rn} \textbf{Hierarchical clustering with of the cuisine space ingredient vectors.} Here, the dendrogram is clustered by the cuisine contribution profiles of ingredients. All leaves from the dendrogram are labelled. The shown clusters arise from the dendrogram cut at $0.3\times$ the dendrogram height, ranked by size. }
\end{figure*}

\subsection{Gene regulatory network}
All genes in the network have associated Gene Ontology (GO) terms, that specifies their biological characteristic or role. There may be multiple GO terms associated with a single gene. A transcription factor (TF) could be regulated upstream by a TF, and regulate a downstream gene. These relationships are specified as incoming or outgoing edges respectively. The GO terms of the in-neighbours and out-neighbours are aggregated to give a GO term profile for each TF, whose contributions are normalised. 

\begin{figure}
\includegraphics[width=0.2\textwidth]{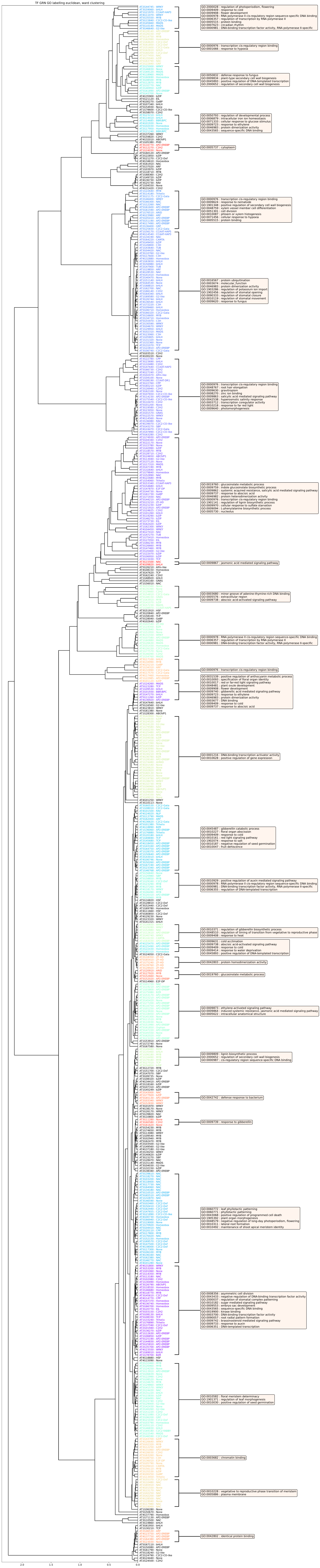}
\centering
\caption{\label{fig:s1:grn}\textbf{Hierarchical clustering of the gene regulatory network vectors.} The leaves are labelled by gene name and transcription factor family annotation. We highlight all enriched clusters ranked from most enriched to least enriched. The GO terms associated to each enriched cluster are shown in the boxes. }
\end{figure}

In figure \ref{fig:s1:grn}, we have applied hierarchical clustering with the Euclidean metric and Ward's method to the TFs, clustering by the neighbourhood GO term profiles. Using the GO term enrichment analysis we extract clusters of TFs in the dendrogram by asking how statistically significant is drawing a group of TFs with a particular set of GO terms. We highlight all the statistically significant GO terms for each enriched cluster, which are sorted by the corrected $p$-value after the Benjamini-Hochberg false discovery rate procedure. The full Gene Ontology (GO) term enriched clusters from the hierarchical clustering are available as a \texttt{.csv} file. 

\section{A further statistical test in the food web network}
For the food web network, we demonstrate a further statistical tests: using the silhouette score for the observed, and the null distribution data. All analysis was done with the vectors as constructed from the finest organism type labelling. The null model is built from randomly shuffling the organism type and recalculating the vectors. 

\begin{figure}
\includegraphics[width=0.4\textwidth]{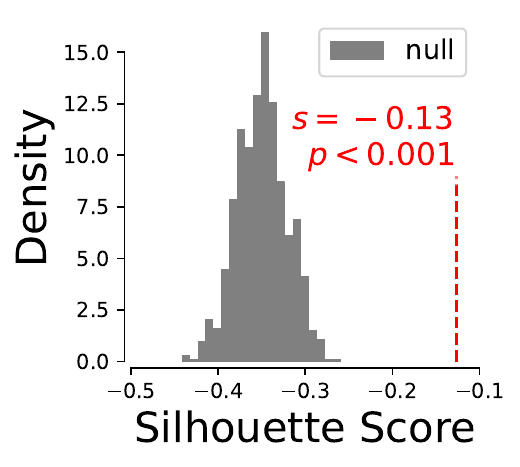}
\centering
\caption{\label{fig:s4:fw_nulls} \textbf{Further null model tests for the food web network.} Here, we shuffle the annotations of the organism types and recalculated the vectors for 1000 realisations to reveal the distribution of silhouette scores (in red $s$) for the identity consistent null model using the finest organism type. See Methods for details. }
\end{figure}

We test against the finest organism type itself (figure \ref{fig:s4:fw_nulls}) and find the silhouette score is negative ($s=-0.13$ in red) which indicates that the embedding is not well aligned with the finest organism type. Additionally, this result is statistically significant ($p<0.001$), and all instances of the random shuffling showed greater magnitude of disagreement between the embedding and the labels. Since we are reassigning the identity of the nodes in each silhouette score calculation, this is an identity consistent null model. 

\end{document}